\documentclass[twocolumn,twoside,slac_two]{revtex4}
\usepackage{graphicx}
\usepackage{fancyhdr}
\begin{document}

\title{Charm Production at RHIC}

%

\author{A. G. Knospe}
\affiliation{Physics Department, Yale University, New Haven, CT
06520, USA}

\begin{abstract}
Observations by the PHENIX and STAR collaborations suggest that a
strongly coupled quark-gluon plasma is produced in heavy-ion
collisions at the Relativistic Heavy Ion Collider (RHIC).  After a
brief introduction to heavy-ion physics, measurements of heavy-quark
production in heavy-ion collisions and the modification of
heavy-quark spectra by the QGP are presented.  Measurements of the
total charm cross-section in several different collision systems
confirm that $c\bar{c}$ pairs are produced through parton
hard-scattering in the initial stages of the collisions.
 Non-photonic $e^{\pm}$ (proxies for heavy quarks) are suppressed by
a factor of $\sim5$ in central Au + Au collisions relative to $p+p$
collisions. This is larger than most current theoretical predictions
and has led to a re-examination of heavy-quark energy loss in the
medium.  The relative contributions of $c$ and $b$ decays to the
non-photonic $e^{\pm}$ spectrum have been predicted by perturbative
QCD calculations; STAR measurements of azimuthal correlation
functions of non-photonic $e^{\pm}$ and hadrons agree with these
predictions.
\end{abstract}

\maketitle

\thispagestyle{fancy}


\section{Introduction}
\begin{figure}[h]
\centering
\includegraphics[width=80mm]{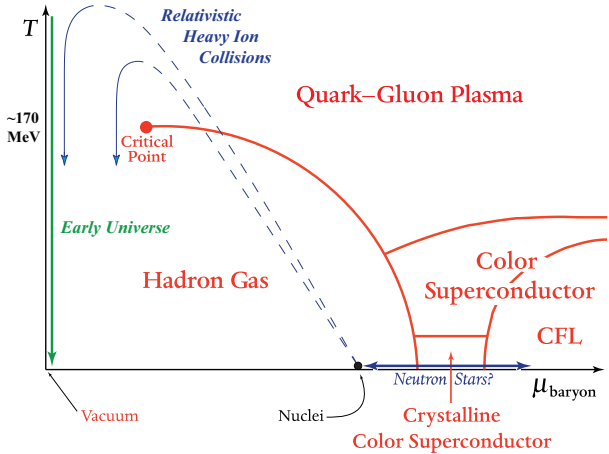}
\caption{Schematic phase diagram of strongly interacting matter. On
the vertical axis is temperature and on the horizontal axis is
baryon chemical potential, $\mu_{baryon}$.} \label{phase_figure}
\end{figure}

The main purpose of the PHENIX and STAR experiments at the
Relativistic Heavy Ion Collider is to study the properties of
strongly interacting matter at high temperatures.  Figure
\ref{phase_figure} shows a schematic theoretical phase diagram of
nuclear matter for various temperatures and baryon chemical
potentials \cite{karsch,kanaya}.  At normal baryon chemical
potentials and low temperatures, strongly interacting matter exists
as nuclei or a gas of interacting hadrons.  At higher temperatures,
however, the quarks may become deconfined due to asymptotic freedom;
the degrees of freedom of the system are not hadrons, but individual
quarks and gluons.  This state of matter, called the quark-gluon
plasma (QGP), is thought to have existed in the first few
microseconds following the Big Bang, and may also have been produced
in high-energy nucleus-nucleus collisions such as those at RHIC at
Brookhaven National Laboratory and the SPS at CERN.

Lattice QCD calculations predict a sudden increase in the number of
degrees of freedom of strongly interacting matter near a critical
temperature $T_{C}\sim170$ MeV \cite{karsch}.  Whether or not there
is a true phase transition and its exact nature has yet to be
determined.  It has been proposed that a critical point may exist in
the phase diagram, and that for higher baryon densities, a
first-order phase transition may be observed \cite{karsch,kanaya}.
A search for this critical point may be conducted in the next few
years at RHIC by reducing the collision CM energy to lower values
($\sim$5GeV) than the collider's present operating range (22 GeV to
200 GeV).

The collision of two nuclei introduces a large amount of energy into
a region of space approximately the size of a nucleus for a short
period of time.  Lattice calculations predict a critical energy
density of $\approx700$ MeV/fm$^{3}$ needed for QGP formation
\cite{karsch}.  The highest-energy RHIC collisions reach an energy
density of at least 4.9 GeV/fm$^{3}$, seven times the critical
energy density \cite{star_energy}.  The QCD vacuum is "melted" and a
quark-gluon plasma is produced.  It is believed that the QGP quickly
reaches thermal equilibrium, expands, and cools for a few fm/$c$
until the transition temperature is reached.  At this point, the
partons become confined into hadrons. The hadron gas expands and the
hadrons scatter inelastically until chemical freeze-out.  The hadron
gas continues to expand for a few more fm/$c$ with elastic
hadron-hadron interactions until thermal freeze-out, after which
hadronic interactions are negligible \cite{blaziot}.  The resulting
shower of particles can be detected in detector systems at RHIC.

Many phenomena in heavy-ion physics depend upon the degree of
overlap between the two colliding nuclei, called the centrality of
the collision.  If the distance between the centers of the nuclei
(impact parameter) is small, the overlap between the nuclei is
large.  Such collisions are called "central" events.  A peripheral
event has a small overlap and the impact parameter approaches the
sum of the radii of the two nuclei.  $N_{part}$ is the number of
nucleons that participate in collisions and $N_{binary}$ is the
number of binary collisions between those participating nucleons.
 $N_{part}$ and $N_{binary}$ are large for central events and small
for peripheral events.  The centrality of an event is estimated
using the multiplicity of charged tracks at mid-rapidity.  Central
collisions are characterized by a higher charged-particle
multiplicity than peripheral events.  $N_{part}$ and $N_{binary}$
are estimated using the Glauber model of nucleus-nucleus
interactions \cite{glauber,star_glauber}.

The ratios of various particle yields have been fit with statistical
thermal models, which indicate that at the time of chemical
freeze-out, the system is thermalized and has a temperature
$T=170-180$ MeV for RHIC collision energies \cite{braun_0}.  A
non-central collision will have a spatially asymmetric overlap
region, roughly elliptically shaped.  The spatial asymmetry
translates into a momentum-space asymmetry in the spectra of final
state particles, with hadrons emitted preferentially near the
reaction plane (the plane containing the beam axis and the impact
parameter vector).  Measurements of this phenomenon (called elliptic
flow) are well described by ideal hydrodynamical models, indicating
that the state of matter produced in the collision has a very low
viscosity and is a nearly perfect liquid
\cite{star_flow_0,star_flow_1,star_flow_2}.  All RHIC experiments
have observed the production of back-to-back di-jets in $p+p$, $d+$
Au, and peripheral Au + Au collisions.  In central Au + Au
collisions, however, di-jets are not observed: the away-side jets
appear to have been quenched by the medium
\cite{star_jets_0,star_jets_1}.

\begin{figure}[h]
\centering
\includegraphics[width=80mm]{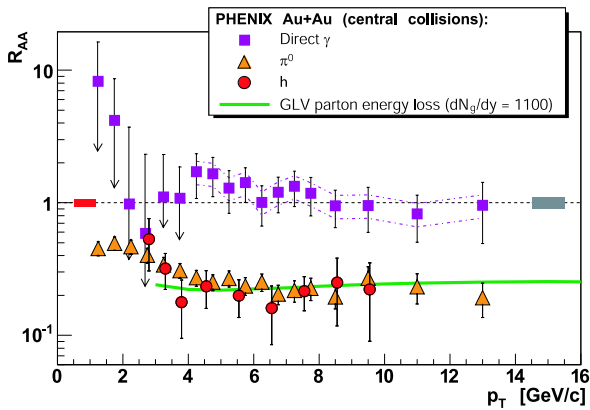}
\caption{Nuclear Modification Factor, $R_{AA}$, for light-flavor
hadrons and direct photons.  $\pi^{0}$ and $\eta$ are suppressed by
a factor of $\sim5$ in central Au + Au collisions at
$\sqrt{s_{NN}}=200$ GeV relative to $p+p$ collisions at the same
energy. Direct photons are not suppressed.  The curve was generated
using a theoretical model of parton interactions with the QGP.  The
data are used to constrain the value of the gluon density
$(dN_{g}/dy)$ in that model.} \label{light_RAA_figure}
\end{figure}

Hard-scattering processes (those with large momentum transfer)
between the incoming nucleons are expected to account for 30-50$\%$
of the particles produced in a nucleus-nucleus collision. Hard
scattering often produces particles with high transverse momentum
($p_{T}>2$ GeV/$c$).  Partons passing through a quark-gluon plasma
are expected to lose energy to the medium through gluon radiation
and collisions with other partons in the medium
\cite{mustafa,djordjevic_0}. The production of a QGP in a
nucleus-nucleus collision will therefore cause a depletion of
high-$p_{T}$ particles relative to proton-proton collisions, which
do not produce a medium. The standard measure of the effect of the
medium on particle yields is the nuclear modification factor
$R_{AA}$.  $R_{AA}$ is the ratio of a particle yield in
nucleus-nucleus collisions to the yield produced in $p+p$
collisions.  The ratio is scaled by $1/\langle N_{binary}\rangle$,
where $\langle N_{binary}\rangle$ is the average number of
nucleon-nucleon collisions in a nucleus-nucleus collision.
\begin{equation} R_{AA}\equiv \frac{\frac{d^{2}N(A+A)}{dp_{T}dy}}{\langle N_{binary}\rangle\frac{d^{2}N(p+p)}{dp_{T}dy}} \label{eq-RAA}
\end{equation}
If no medium is produced, then a nucleus-nucleus collision can be
viewed as an incoherent superposition of nucleon-nucleon collisions
and $R_{AA}$ will be unity. Deviations from unity indicate the
effects of nuclear matter and the quark-gluon plasma on particle
yields. Central collisions produce a larger medium than peripheral
collisions, which should result in a greater suppression of
high-$p_{T}$ particles.

Figure \ref{light_RAA_figure} shows measurements of $R_{AA}$ for
direct photons, $\pi^{0}$, and $\eta$ by PHENIX.  $R_{AA}$ is unity
for direct photons \cite{phenix_gamma}, indicating that there is no
suppression of direct photons in nucleus-nucleus collisions. This is
expected since photons do not interact strongly and should not be
affected by the presence of a QGP.  However, the hadrons
\cite{phenix_pi0_0,phenix_pi0_1,phenix_eta} are suppressed by about
a factor of 5 at high $p_{T}$. Also shown is a theoretical
prediction of light-flavor-hadron suppression from the GLV model of
parton-QGP interactions \cite{vitev_0,vitev_1}. $dN_{g}/dy$ is the
gluon density, a parameter in the GLV model related to the opacity
of the medium. The observed light-flavor-hadron suppression was used
to constrain the value of this parameter, giving
$dN_{g}/dy\approx1100$. $R_{AA}$ has also been measured for
heavy-flavor decay products and compared to models; this will be
discussed in subsequent sections. These and other measurements
indicate that the matter produced in RHIC collisions is a strongly
coupled quark-gluon plasma (sQGP) \cite{gyulassy_0}.

\section{Experiments}
The experiments described in these proceedings are conducted at the
Relativistic Heavy Ion Collider (RHIC) at Brookhaven National
Laboratory on Long Island, NY \cite{rhic}.  Ions are accelerated
from the tandem Van de Graaff accelerators or the proton sources,
through the AGS Booster and the AGS, and are injected into RHIC.
Electrons are stripped off at several locations along the way.  RHIC
consists of two synchrotron rings, 3.8 km in circumference.  The two
counter-circulating beams of ions intersect in each of the six
interaction regions.  The PHENIX and STAR detectors sit at two of
these interaction regions.  At RHIC, physicists can create and
maintain beams of ions ranging from protons (both polarized and
unpolarized) to the heaviest nuclei.  RHIC can be used to collide
protons with a center-of-mass collision energy up to $\sqrt{s}=500$
GeV and a luminosity of $\sim10^{32}$ cm$^{-2}$s$^{-1}$. RHIC can
also collide ions ranging in mass from deuterons to gold nuclei with
CM energy per nucleon pair $\sqrt{s_{NN}}\leq200$ GeV and a
luminosity of $\sim10^{26}$ cm$^{-2}$s$^{-1}$ \cite{star_detector}.

\begin{figure}[h]
\centering
\includegraphics[width=80mm]{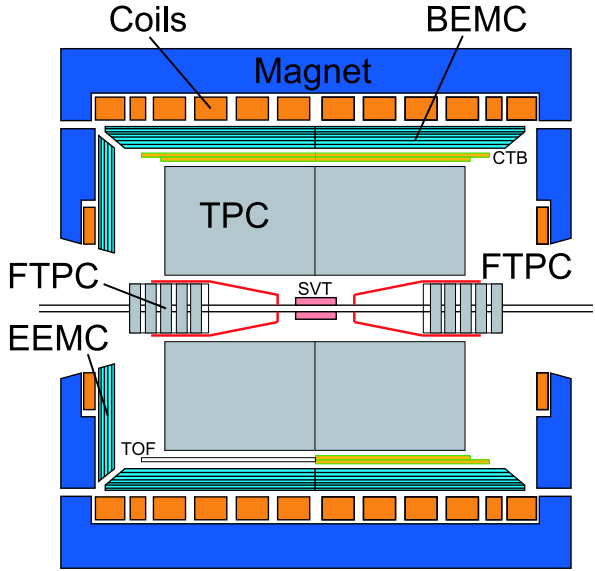}
\caption{The STAR detector.  Measurements described below use the
Time Projection Chamber (TPC), Time-of-Flight detector (TOF), and
the electromagnetic calorimeter (EMC).} \label{STAR_figure}
\end{figure}
\begin{figure}[h]
\centering
\includegraphics[width=80mm]{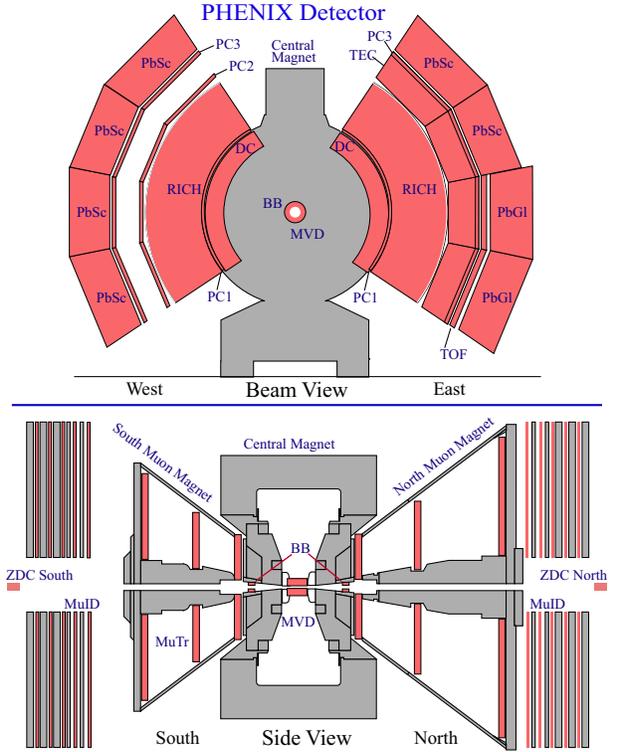}
\caption{The PHENIX detector in 2003.  Measurements described below
use the Pad Chambers (PC), Drift Chambers (DC), Time Expansion
Chambers (TEC), Ring-Imaging Cherenkov detectors (RICH),
Time-of-Flight detector (TOF), and the lead-scintillator (PbSc) and
lead-glass (PbGl) electronmagnetic calorimeters.}
\label{PHENIX_figure}
\end{figure}

Figure \ref{STAR_figure} shows a diagram of the STAR
(\textbf{S}olenoidal \textbf{T}racker \textbf{A}t \textbf{R}HIC)
detector \cite{star_detector}.  The barrel and forward
time-projection chambers (TPC and FTPC) record particle trajectories
inside a room-temperature solenoidal 0.5-T magnet.  The TPC covers
pseudorapidity $|\eta|<1.8$, while the FTPCs cover $2.5<|\eta|<4$.
After passing through the TPC, particles enter the barrel
electromagnetic calorimeter (BEMC) or the endcap electromagnetic
calorimeter (EEMC, not shown).  The STAR calorimeters together cover
pseudorapidity -$1<\eta<2$.  The Shower Maximum Detectors (SMDs),
located approximately 5 radiation lengths inside each EMC tower
module, provide additional particle identification based on the
shape of the electromagnetic shower produced in the calorimeters.
 The SMDs allow shower shapes to be measured to high precision
($\Delta\eta=0.007$, $\Delta\phi=0.007$ rad).  A silicon vertex
tracker (SVT) covers $|\eta|<1$ between the beampipe and the TPC,
providing accurate particle tracking near the collision vertex. A
prototype Time-of-Flight detector (TOF) has been installed outside
the TPC covering $-1<\eta<0$ and $\Delta\phi=0.04$ rad.  The TOF
provides a precise measurement of particle velocity. Plans call for
the TOF to be extended to full azimuthal coverage over
pseudorapidity $|\eta|<1$.  A full description of the STAR detector
is given in \cite{star_detector}.

Figure \ref{PHENIX_figure} shows a diagram of the PHENIX
(\textbf{P}ioneering \textbf{H}igh-\textbf{E}nergy \textbf{N}uclear
\textbf{I}nteraction E\textbf{x}periment) detector.
\cite{phenix_detector} The two central spectrometer arms sit in an
axial magnetic field and cover $|\eta|<0.35$ and $\pi/2$ each in
azimuth. Particle tracking is provided by three sets of pad chambers
and (in the east arm) the time expansion chamber. The Ring-Imaging
Cherenkov detectors and the Time-of-Flight detector provide particle
identification.  Beyond these detector systems sit lead-glass and
lead-scintillator calorimeters.  The two forward muon spectrometer
arms sit in radial magnetic fields.  They consist of drift chambers
for precision tracking and muon identifiers.  The muon identifiers
are made up of alternating layers of steel absorber plates and
streamer-tube tracking layers.  A full description of the PHENIX
detector, including the azimuthal and pseudorapidity coverage of
each detector subsystem, is given in \cite{phenix_detector}.

\section{Heavy Flavors}
Due to the high luminosity at RHIC, even particles with
comparatively low production cross-sections, such as $c$ and $b$
quarks, can be used to probe the strongly interacting matter
produced.  In nucleus-nucleus collisions, the dominant production
mechanism for heavy quarks is gluon-gluon fusion in the initial hard
scattering of nucleons \cite{z_lin}.  Thermal production of
heavy-quark pairs in the medium is believed to be negligible after
the initial stages of the collision and the number of heavy quarks
is essentially "frozen."  Therefore, heavy quarks are probes
sensitive to all stages in the evolution of the QGP, from its
initial formation to hadronization and freeze-out.

The presence of a QGP is expected to lead to the dissocation and
suppression of heavy quarkonia through Debye screening of color
charges. Some theoretical calculations predict sequential
dissociation of heavy quarkonia, with the more weakly bound
resonances dissociating at lower temperatures \cite{matsui}.  If
this is true, measurements of heavy quarkonium suppression could
lead to a measurement of the initial temperature of the medium.
PHENIX measurements of $R_{AA}$ for $J/\psi$
\cite{phenix_jpsi_0,phenix_jpsi_1,phenix_jpsi_2} show less
suppression than was expected based on SPS data. This may be due to
the regeneration of $J/\psi$ through the recombination of
dissociated charm (anti)quarks into new $c\bar{c}$ pairs
\cite{braun_1,braun_2,grandchamp,thews,andronic}.

While all partons lose energy to the medium through gluon
bremsstrahlung, the intensity and angular distribution of the gluon
radiation is predicted to depend upon the mass of the radiating
parton.  In the small-angle approximation, the angular distribution
of gluon radiation from heavy quarks differs from the bremsstrahlung
spectrum for massless partons by the pre-factor given in equation
\ref{eq-dead_cone}.
\begin{equation} \frac{dP(heavy\;quark)}{d\theta^{2}}=(1+\frac{M_{Q}}{E_{Q}}\cdot\frac{1}{\theta^{2}})^{-2}\frac{dP(light\;quark)}{d\theta^{2}}\label{eq-dead_cone}
\end{equation}
where $M_{Q}$ and $E_{Q}$ are the heavy-quark mass and energy,
respectively \cite{kharzeev}.  This suppression of gluon radiation
from heavy quarks is strongest for angles less than $M_{Q}/E_{Q}$.
The suppression of small-angle gluon radiation is called the dead
cone effect.  Because of the dead cone effect, heavy quarks should
lose less energy to the medium through gluon radiation than light
quarks. Partons can also lose energy through collisions with other
partons in the medium.  For heavy quarks traversing the medium, the
amount of energy lost through collisions may be comparable to the
amount lost through gluon radiation \cite{mustafa,djordjevic_0}.

\section{Measurements of Open Heavy Flavors}
The RHIC experiments have studied several different heavy-flavor
decay channels, including the hadronic decays of $D^{0}$ mesons to
pions and kaons, and the semileptonic decays of heavy-flavor hadrons
to muons and electrons.

\begin{figure*}[t]
\centering
\includegraphics[width=135mm]{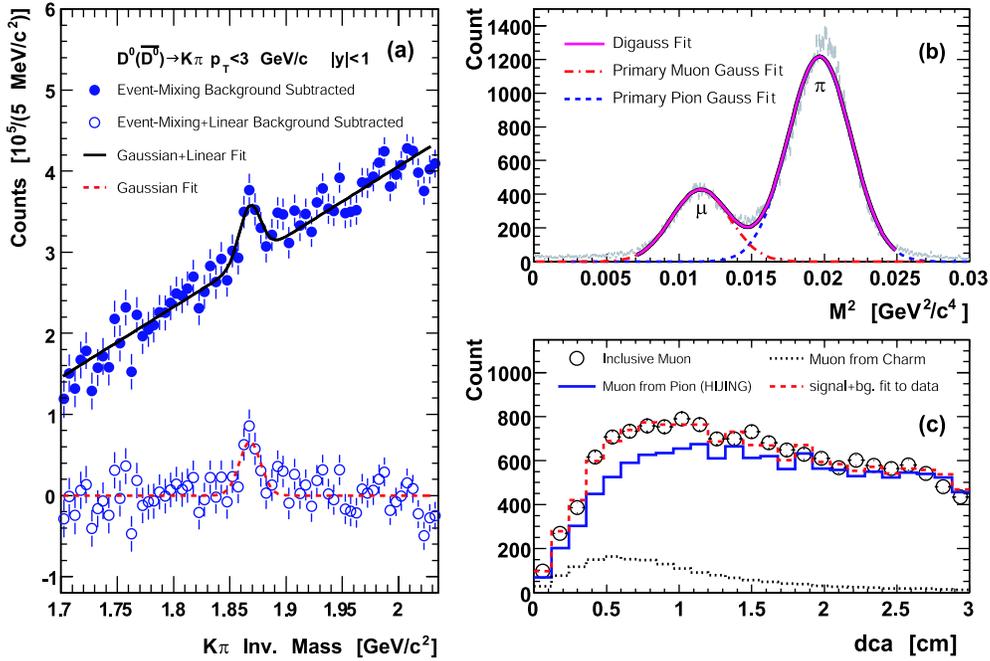}
\caption{STAR heavy-flavor identification methods.  (a) $D^{0}$
meson reconstruction in 200-GeV Au+Au collisions.  The $K\pi$
invariant mass distribution is shown after the subtraction of the
combinitorial background (estimated using event-mixing).  The
residual background is approximated by a linear function and
subtracted.  (b) $m^{2}$ distribution measured by TOF detector in
200-GeV Au + Au collisions. The particles have passed the
energy-loss cut used to identify low-$p_{T}$ muons.  (c) DCA
distributions used to identify charm-decay contribution to the
inclusive muon spectrum (see explanation in text).}
\label{D0_figure}
\end{figure*}

\subsection{Direct Reconstruction of $D^{0}$ Decays}
The STAR collaboration has found the yields of $D^{0}$ and
$\bar{D^{0}}$ mesons \cite{star_charm_dAu} by reconstructing the
$D^{0}(\bar{D}^{0})\rightarrow K^{\mp}+\pi^{\pm}$ decays, which have
a branching ratio of 3.83$\%$ \cite{pdg}.  The pions and kaons are
identified by their energy loss in the Time Projection Chamber. STAR
cannot reconstruct the full decay topology since $c\tau(D^{0}) =
124\;\mu$m \cite{pdg} and the TPC does not have sufficient track
projection resolution to distinguish $D^{0}$ decay products from
tracks coming directly from the primary collision vertex.  The
$D^{0}$ invariant mass spectrum was obtained by pairing each kaon
with oppositely charged pions from the same event.  The
combinatorial background was estimated through event mixing
techniques and subtracted. Figure \ref{D0_figure}a shows the $K\pi$
invariant mass spectrum for $|y|<1$ in Au + Au collisions at
$\sqrt{s_{NN}} = 200$ GeV \cite{zhang}.  A clear $D^{0}$ peak is
visible. Because of the small branching ratio for this decay and the
lack of a dedicated trigger, this analysis is limited by statistics
to $p_{T} < 3$ GeV/$c$.

\subsection{Decays to Muons}
The STAR collaboration has identified low-$p_{T}$ muons (0.17
GeV/$c<p_{T}<0.25$ GeV/$c$) through measurements of energy loss in
the Time Projection Chamber and $m^{2}=(p/(\beta\gamma))^{2}$ in the
Time-of-Flight detector and the TPC.  Figure \ref{D0_figure}b shows
the $m^{2}$ distribution of particles after energy-loss selection
\cite{zhong}.  The muon and pion peaks are clearly visible. In
addition to muons from heavy flavor decays, the muon peak contains a
large number of muons from pion and kaon decays; it was necessary to
remove these "background" muons.  HIJING \cite{hijing} was used to
simulate the DCA (distance of closest approach) of muon tracks with
the primary collision vertex (see Figure \ref{D0_figure}c).  The DCA
distribution for muons from charm decays has a maximum much closer
to zero than the distribution for muons from pion and kaon decays.
The observed DCA distribution was fit with a linear combination of
these two simulated DCA distributions to obtain the contribution of
charm-decay muons to the total muon yield. \cite{zhong}

\begin{figure}[h]
\centering
\includegraphics[width=85mm]{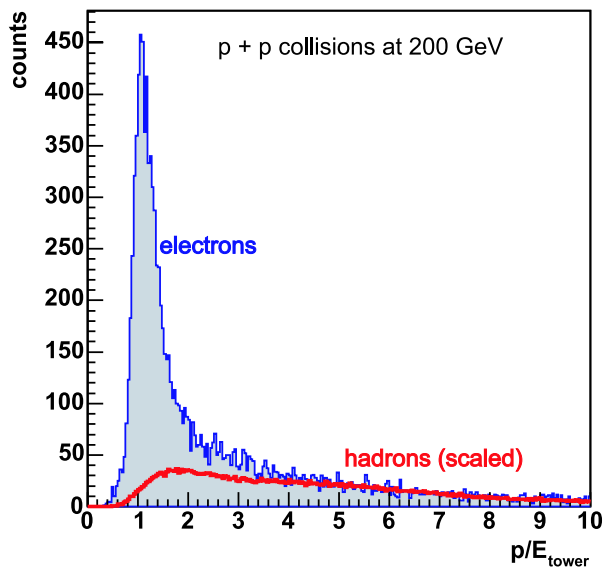}
\caption{$p/E$ distributions for $e^{\pm}$ and hadrons in 200-GeV
$p+p$ collisions at STAR.  $e^{\pm}$ and hadrons are identified
through their energy loss in the Time Projection Chamber. The hadron
distribution has been scaled to match the $e^{\pm}$ distribution at
high values of $p/E$.  This provides an estimate of the hadron
background that remains in the $e^{\pm}$ peak.} \label{poe_figure}
\end{figure}

\subsection{Decays to Single Electrons}
Both the PHENIX and STAR collaborations have also studied the
spectrum of single electrons, i.e. $e^{\pm}$ produced with
(anti)neutrinos in weak decays.  The spectrum of single $e^{\pm}$ is
expected to be dominated by heavy-flavor semileptonic decays (e.g.
$D^{0}\rightarrow e^{+}+K^{-}+\nu_{e}$).  The main sources of
background to the single $e^{\pm}$ signal are $e^{+}e^{-}$ pairs
from $\pi^{0}$ and $\eta$ Dalitz decays and photon conversions
\cite{pdg}. For this reason, single $e^{\pm}$ are called
non-photonic $e^{\pm}$, while the background $e^{\pm}$ are called
photonic $e^{\pm}$. The decays of vector mesons (e.g. $\rho$,
$\phi$, and $\omega$) make small contributions to the photonic
$e^{\pm}$ background. Background $e^{\pm}$ from photon conversion is
less significant at PHENIX due to the reduced amount of material
relative to STAR. $K_{e3}$ decays
($K^{\pm}\rightarrow\pi^{0}+e^{\pm}+\nu_{e}(\bar{\nu}_{e})$ and
$K^{0}_{L}\rightarrow\pi^{\mp}+e^{\pm}+\nu_{e}(\bar{\nu}_{e})$) make
small contributions to the single $e^{\pm}$ signal.

The STAR collaboration identifies $e^{\pm}$ using two different
methods. For $p_{T}<3.5$ GeV/$c$, $e^{\pm}$ are identified through
measurements of energy loss and $m^{2}$ with the Time Projection
Chamber and Time-of-Flight detector (similar to the method of muon
identification described above) \cite{star_charm_dAu}.  For
$p_{T}>1.5$ GeV/$c$, $e^{\pm}$ are identified through measurements
of TPC energy loss, the energy $E$ deposited in the electromagnetic
calorimeter, and the shape of the electromagnetic shower measured in
the shower maximum detector \cite{star_nphe}.  The use of a trigger
allows this measurement to extend up to $p_{T}\approx8$ GeV/$c$. An
energy-loss cut of approximately 3.5 keV/cm $<dE/dx<$ 5 keV/cm is
used to identify $e^{\pm}$ (the exact bounds are varied slightly
depending on track momentum and event charged-track multiplicity).
Compared to hadrons, $e^{\pm}$ produce larger showers and deposit
more of their energy in the EMC. The ratio $p/E$ has a maximum near
1 for $e^{\pm}$ (see Figure \ref{poe_figure} \cite{suaide_0}) and
additional $e^{\pm}$ identification is provided by cuts on the
shower size in the SMD. The remaining hadron contamination is
$\approx2\%$ at $p_{T}\sim2$ GeV/$c$ and $\approx20\%$ at
$p_{T}\sim8$ GeV/$c$. STAR identifies photonic (background)
$e^{\pm}$ through invariant-mass reconstruction of $e^{+}e^{-}$
pairs. The invariant-mass distribution of $e^{+}e^{-}$ pairs from
photon conversions and $\pi^{0}$ and $\eta$ Dalitz decays has a
maximum near 0 mass. When a cut of $M_{inv}(e^{+}e^{-})<150$
MeV/$c^{2}$ is used, photonic electrons are identified with an
efficiency around 70$\%$.

\begin{figure*}[t]
\centering
\includegraphics[width=170mm]{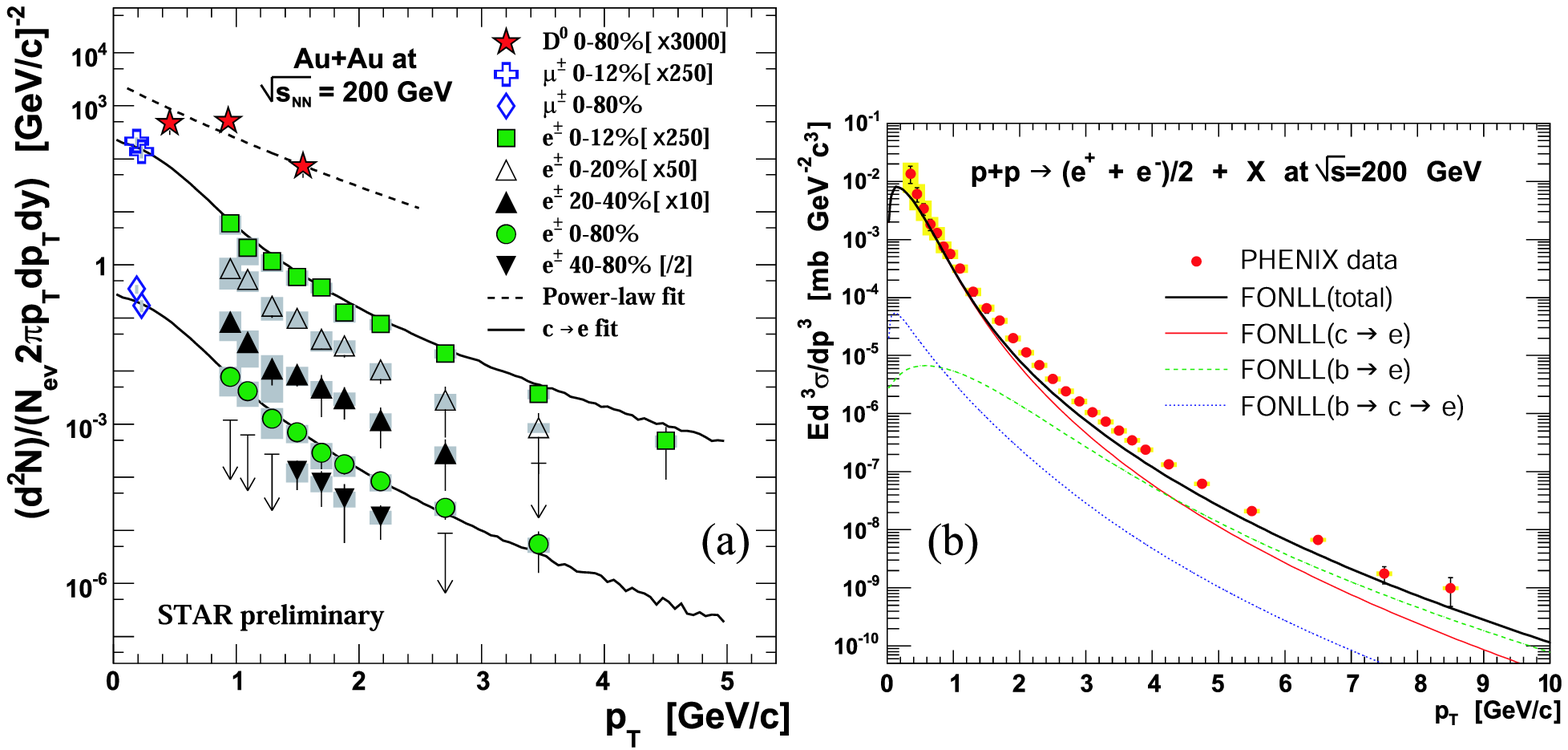}
\caption{Measurements used in calculation of total charm
cross-section.  (a) STAR $D^{0}$, muon, and $e^{\pm}$ spectra for
200-GeV Au + Au collisions in various centrality bins.  (b) PHENIX
non-photonic $e^{\pm}$ spectrum for 200-GeV $p+p$ collisions.  Also
shown are pQCD predictions (FONLL) for the $c$- and $b$-decay
contributions to the spectrum (see discussion below and Figure
\ref{fonll_figure}).} \label{xsection_spectra_figure}
\end{figure*}

The PHENIX collaboration identifies $e^{\pm}$ using the Ring-Imaging
Cherenkov detector, measurements of the shower shape in the
electromagnetic calorimeter, and a cut on the energy-to-momentum
ratio \cite{phenix_nphe,phenix_nphe_AuAu}. Photonic $e^{\pm}$ are
identified using two different methods. In the "cocktail
subtraction" method, the spectra of $e^{\pm}$ from various sources
of background are simulated. Measured yields of $\pi^{0}$, $\eta$,
direct photons, and other sources of background are used as input
for the simulation generator.  In the "converter subtraction"
method, a photon converter (a thin brass sheet of 1.67$\%\;X_{0}$)
is inserted around the beam pipe.  $\Delta N_{e}$, the increase in
the $e^{\pm}$ yield due to the converter, is measured. GEANT
simulations are used to determine $R_{\gamma}$, the fractional
increase in the photonic $e^{\pm}$ yield caused by the converter
($R_{\gamma}\approx2.3$ in the 2006 $p+p$ run). Knowledge of $\Delta
N_{e}$ and $R_{\gamma}$ allows the photonic $e^{\pm}$ yield to be
determined and removed from the inclusive $e^{\pm}$ yield.  Where
$N^{NC}_{\gamma e}$ is the photonic $e^{\pm}$ yield with no
converter present,\begin{equation}N^{NC}_{\gamma e}=\frac{\Delta
N_{e}}{R_{\gamma}-1}.\label{eq-converter}
\end{equation} The non-photonic $e^{\pm}$ yields measured
using the cocktail and converter subtraction methods are consistent
with each other.

\begin{figure*}[t]
\centering
\includegraphics[width=170mm]{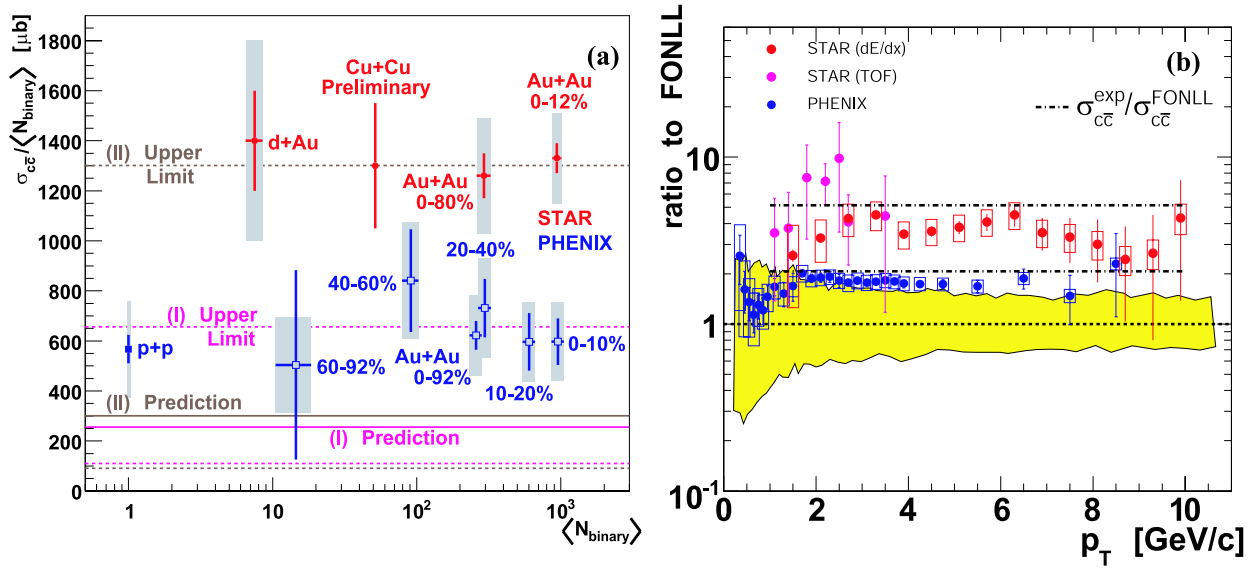}
\caption{Comparisons of RHIC measurements to FONLL predictions. (a)
Total charm cross-section (scaled by $1/\langle N_{binary}\rangle$)
in different collision systems at $\sqrt{s_{NN}}=200$ GeV.
Theoretical predictions are also shown.  (b) Ratios of non-photonic
$e^{\pm}$ yields in central 200-GeV Au + Au collisions for STAR and
PHENIX to FONLL prediction (I) in central 200-GeV Au + Au
collisions.  STAR $e^{\pm}$ have been selected using measurements of
energy-loss ($dE/dx$) in the TPC or $m^{2}$ in the TOF.}
\label{compare_fonll_figure}
\end{figure*}

\section{Total Charm Cross-Section}
STAR determines the total charm cross-section, $\sigma_{c\bar{c}}$,
for each collision system through a combined fit of the $D^{0}$,
muon, and non-photonic $e^{\pm}$ measurements described above (e.g.
Figure \ref{xsection_spectra_figure}a for 200-GeV Au + Au
collisions) \cite{zhong}. PHENIX determines the total charm
cross-section from the measurement of the non-photonic $e^{\pm}$
yield (e.g. Figure \ref{xsection_spectra_figure}b for 200-GeV $p+p$
collisions) \cite{phenix_nphe}. Figure \ref{compare_fonll_figure}a
shows measurements of the scaled total charm cross-section by STAR
and PHENIX for $p+p$ \cite{phenix_nphe}, $d$ + Au
\cite{star_charm_dAu}, and Au + Au \cite{phenix_nphe_AuAu,star_nphe}
collisions (in different centrality bins) at $\sqrt{s_{NN}} = 200$
GeV.  A preliminary STAR measurement for Cu + Cu using only $D^{0}$
reconstruction is also shown \cite{baumgart}.  The cross-section is
divided by $\langle N_{binary}\rangle$, the average number of binary
collisions for the given collision system.  Within each experiment,
the charm cross-section scales with the number of binary collisions.
This is a confirmation that charm is indeed produced through initial
parton hard-scattering and that charm production through other
mechanisms (such as thermal production in the QGP) is not
significant.  Note that the cross-sections measured by STAR are
higher than those measured by PHENIX by a factor of $\sim$2.  This
is still under investigation.

Due to the large quark masses, the hard-scattering processes that
produce heavy flavor can be calculated using perturbative QCD
\cite{frixone}.  The most advanced perturbative calculation scheme
is the Fixed-Order plus Next-to-Leading-Log-resummed approximation,
or FONLL.  A FONLL prediction for the charm cross-section
\cite{cacciari,vogt} is shown as prediction (I) in Figure
\ref{compare_fonll_figure}a. The PHENIX data are consistent with
this prediction;  the STAR data are greater than the FONLL
prediction by a factor of $\sim$5 and sit well above the upper
uncertainty bound.  However, the charm cross-section predictions are
sensitive to the number of active flavors, the choice of scale, and
the parton densities in the collision system. A new calculation by
R. Vogt \cite{vogt} (prediction (II) in Figure
\ref{compare_fonll_figure}a) indicates that the uncertainties on the
perturbative calculation may be larger than previously thought.  The
total charm cross-section measured by STAR is consistent with the
new perturbative calculation.

Figure \ref{compare_fonll_figure}b shows the ratios of the measured
non-photonic $e^{\pm}$ yields
\cite{star_nphe,star_charm_dAu,phenix_nphe} to the FONLL prediction
as functions of $p_{T}$ for $p+p$ collisions at $\sqrt{s} = 200$
GeV. The dashed horizontal lines indicate the ratio of the measured
cross-sections to the FONLL cross-section. Figure
\ref{compare_fonll_figure}b indicates that STAR's disagreement with
PHENIX (by a factor of $\sim$2) and FONLL (by a factor of $\sim$5)
exists even at high $p_{T}$. However, the FONLL prediction does
describe the shape of the STAR and PHENIX non-photonic $e^{\pm}$
spectra well.

\begin{figure*}[t]
\centering
\includegraphics[width=170mm]{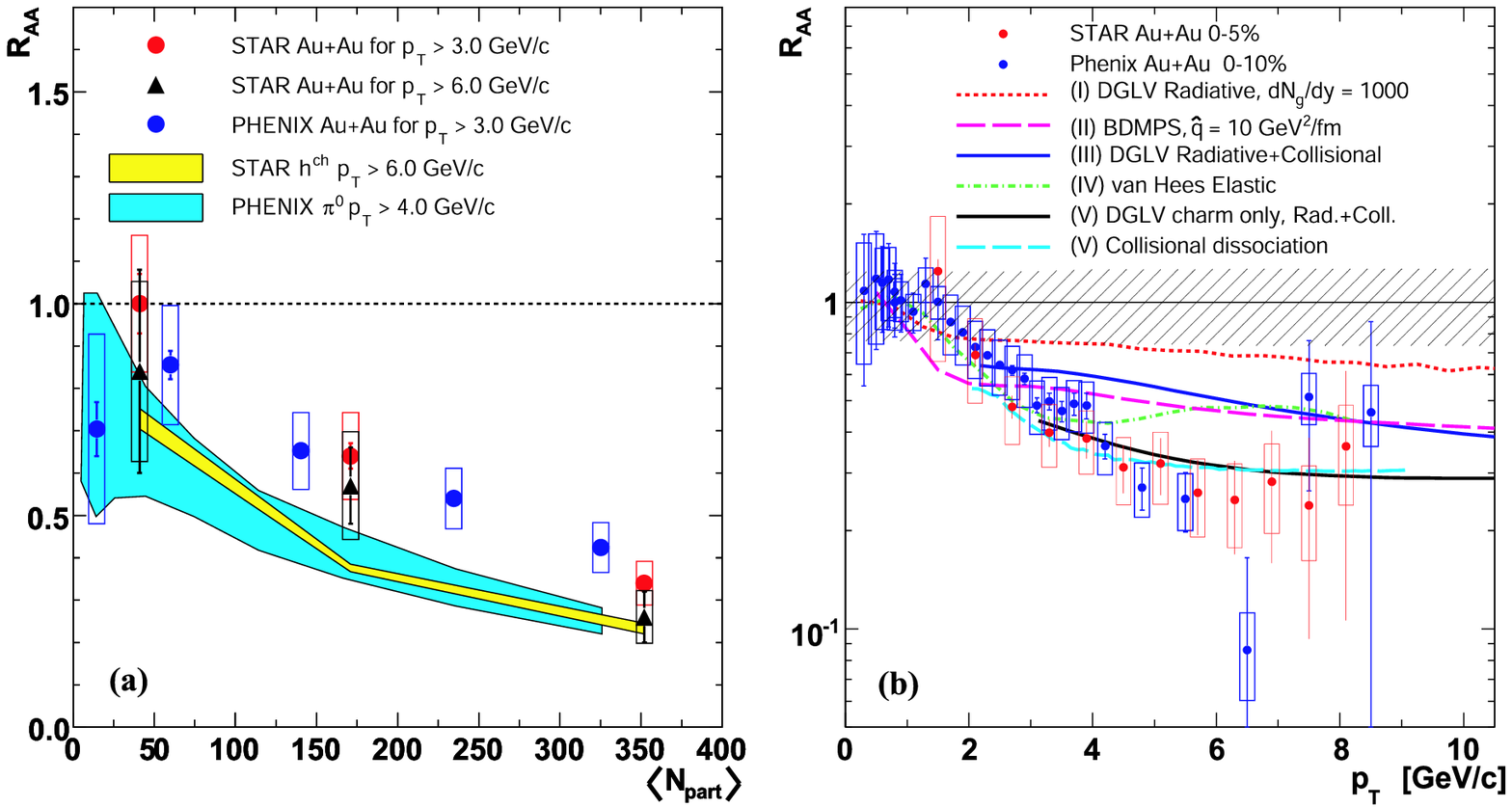}
\caption{Nuclear modification factor $(R_{AA})$ for non-photonic
$e^{\pm}$ in Au + Au collisions at 200 GeV. (a) $R_{AA}$ vs.
$\langle N_{part}\rangle$.  Also shown are light-flavor hadron
$R_{AA}$ data. (b) $R_{AA}$ vs. $p_{T}$ for central collisions.  The
data are compared to theoretical predictions.} \label{RAA_figure}
\end{figure*}

\section{Medium Modification of Non-photonic $e^{\pm}$ Spectra}

The cross-section discrepancy between PHENIX and STAR cancels in the
nuclear modification factor $R_{AA}$, which is the scaled ratio of
particle yields (see definition above).  Figure \ref{RAA_figure}a
shows $R_{AA}$ of non-photonic $e^{\pm}$ for Au + Au collisions at
$\sqrt{s_{NN}} = 200$ GeV for both PHENIX \cite{phenix_nphe_AuAu}
and STAR \cite{star_nphe,suaide_1}.  $R_{AA}$ is plotted as a
function of $\langle N_{part}\rangle$, the average number of
nucleons participating in collisions in a given centrality bin.  The
most central collisions have the highest values of $\langle
N_{part}\rangle$. The PHENIX and STAR $R_{AA}$ are consistent with
each other across the range of $\langle N_{part}\rangle$ shown.
Plotted in shaded bands are the values of $R_{AA}$ measured by
PHENIX for $\pi^{0}$ \cite{phenix_pi0_0} and by STAR for charged
light-flavor hadrons \cite{star_light}. The measured suppression of
non-photonic $e^{\pm}$ is similar to the suppression observed for
light-flavor hadrons. This was unexpected, as heavy quarks were
expected to lose less energy in the medium than light quarks.  As a
result, non-photonic $e^{\pm}$ (proxies for heavy quarks) would be
suppressed less than light-flavor hadrons.

Figure \ref{RAA_figure}b shows a comparison of the non-photonic
$e^{\pm}$ suppression measured by STAR and PHENIX to several
theoretical models of heavy-quark interactions with a quark-gluon
plasma.  The data are from central Au + Au collisions at
$\sqrt{s_{NN}}=200$ GeV.  In $d$ + Au collisions at this energy a
$\sim50\%$ \textit{enhancement} of non-photonic $e^{\pm}$ is
observed (data \cite{star_nphe} not shown).  The enhancement is
explained by the Cronin effect (multiple scattering in normal
nuclear matter). Therefore, the strong non-phonic-$e^{\pm}$
suppression observed in central Au + Au collisions does not appear
to be due to such cold nuclear matter effects.  The calculations
that produce curves (I) and (II) include only energy loss through
gluon radiation; these calculations under-predict the suppression of
non-photonic $e^{\pm}$ \cite{djordjevic_1,armesto}.  $dN_{g}/dy$ is
the gluon density in the DGLV model (curve (I)) and $\hat{q}$ is the
time-averaged transport coefficient in the BDMPS model (curve (II)).
These parameters are related to the opacity of the QGP, with higher
values corresponding to a more opaque medium.  Light-hadron
suppression data are used to constrain the values of these
parameters.  The DGLV model with $dN_{g}/dy\approx1000$ and the
BDMPS model with $\hat{q}\approx10$ GeV$^{2}$/fm describe the
observed light-hadron suppression well.

In addition to energy loss through gluon radiation, partons can lose
energy through collisions with other partons in the medium.
Calculations \cite{mustafa} indicate that, for heavy quarks,
collisional energy loss is as important as radiative energy loss.
Curve (III) is generated using the DGLV model (which produced curve
(I)) including both radiative and collisional energy loss
\cite{wicks}. Curve (IV) is generated using the model of van Hees
\textit{et al}. \cite{van_hees}, which includes heavy-quark energy
loss through elastic collsions in the medium and the formation of
resonant $D$- and $B$-meson states through quark coalescence. Curves
(III) and (IV) seem to describe the data better than curves (I) and
(II), which include only radiative energy loss, but still tend to
under-predict the observed suppression of non-photonic $e^{\pm}$,
especially at high $p_{T}$.  The model of Adil and Vitev \cite{adil}
(curve (VI)) uses the fragmentation of heavy quarks and the
subsequent dissociation of $D$- and $B$-meson states in the medium
to explain the suppression pattern of non-photonic $e^{\pm}$.  This
model does describe the observed suppression well.

Curves (I)-(IV) and (VI) were generated using FONLL predictions
\cite{cacciari} for the relative contributions of $c$- and $b$-
quark decays to the non-photonic $e^{\pm}$ spectrum.  Curve (V) is
generated with the assumption that only $c$-quark decays contribute
to the non-photonic $e^{\pm}$ spectrum and that the $b$-quark
contribution is insignificant for the $p_{T}$ range shown
\cite{wicks}. Curve (V) (which includes both radiative and
collisional energy loss) is consistent with the measured
suppression.  It is therefore important to disentangle the $c$ and
$b$ contributions and determine their relative strengths; this will
be discussed in the next section.

\section{Electron-Hadron Azimuthal Correlations}
\begin{figure}[h]
\centering
\includegraphics[width=80mm]{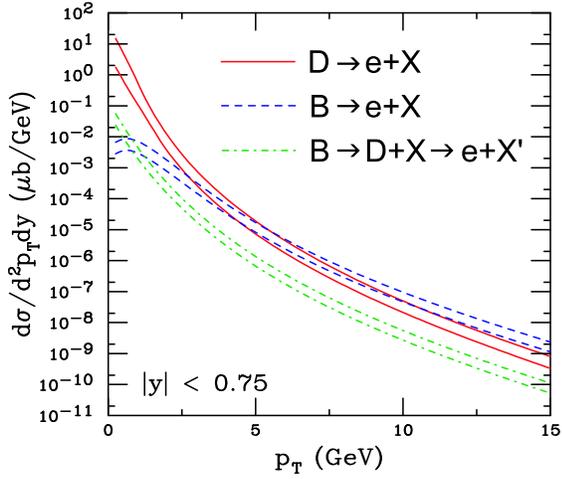}
\caption{FONLL predictions for the contributions of $D$- and
$B$-meson decays to the mid-rapidity non-photonic $e^{\pm}$
cross-section at $\sqrt{s}=200$ GeV.} \label{fonll_figure}
\end{figure}

Figure \ref{fonll_figure} shows FONLL calculations of the $D$- and
$D$-meson-decay contributions to the non-photonic electron
cross-section as functions of $p_{T}$ \cite{cacciari}. The upper and
lower limits result from the variation of model parameters including
quark masses and renormalization scales. The charm contribution is
dominant at low $p_{T}$.  The bottom contribution becomes larger
than the charm contribution around $p_{T}\sim5$ GeV/$c$.

\begin{figure}[h]
\centering
\includegraphics[width=80mm]{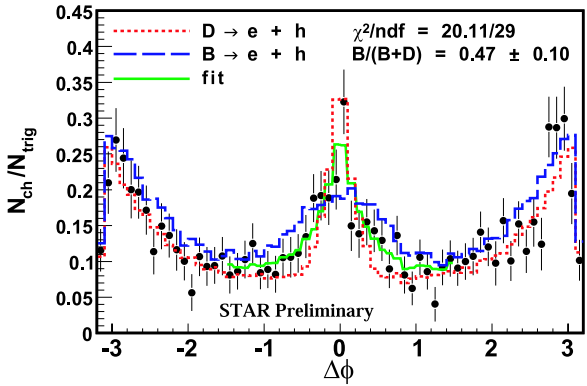}
\caption{$\Delta\phi$ for pairs of non-photonic $e^{\pm}$ and
hadrons. 5.5 GeV/$c< p_{T}(e^{\pm}) < 6.5$ GeV/$c$ and $p_{T}(h) >
0.3$ GeV/$c$. On the vertical axis is the number of charged tracks
per trigger $e^{\pm}$.} \label{eh_dist_figure}
\end{figure}
\begin{figure}[h]
\centering
\includegraphics[width=80mm]{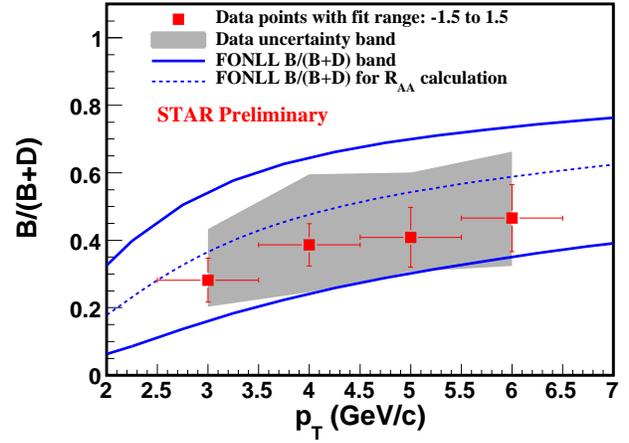}
\caption{Fractional contribution of $B$-meson decays to the
non-photonic $e^{\pm}$ spectrum at $\sqrt{s}=200$ GeV.}
\label{eh_fraction_figure}
\end{figure}

To disentangle the charm and bottom contributions, the STAR
collaboration has studied azimuthal correlations between
non-photonic $e^{\pm}$ and hadrons.  Figure \ref{eh_dist_figure}
shows the difference in azimuthal angle $(\Delta\phi)$ between
non-photonic $e^{\pm}$ and hadrons in 200-GeV proton-proton
collisions for one $p_{T}$ bin \cite{x_lin}.  Due to decay
kinematics, there will be an azimuthal correlation between the
leptons and hadrons produced in heavy-flavor semi-leptonic decays.
Due to the larger $B$-meson mass, a $B$-meson can give more kinetic
energy to its decay products than a $D$-meson, resulting in a
broader near-side ($\Delta\phi\sim0$) correlation peak.  The
expected charm and bottom contributions to this distribution are
simulated with PYTHIA. Varying the quark fragmentation functions
inside PYTHIA does not significantly alter the shapes of the
simulated $\Delta\phi$ distributions.  The measured distribution is
then fit with a linear combination of the simulated charm and bottom
distributions to find their relative strengths.  $B/(B+D)$, the
fraction of the total non-photonic $e^{\pm}$ cross-section due to
$B$-meson decays, is plotted in Figure \ref{eh_fraction_figure}.
The measurement of $B/(B+D)$ through $e$-$h$ correlations
\cite{x_lin} is consistent with FONLL predictions. This is an
indication that $b$-quark suppression \textit{should} be taken into
account in describing the suppression of non-photonic $e^{\pm}$ at
moderate $p_{T}$ (see the discussion in the previous section).

\section{Conclusions}
Measurements of jet quenching and elliptic flow suggest that the
state of matter created in a heavy-ion collision at RHIC energies is
a strongly coupled quark-gluon plasma (sQGP) \cite{gyulassy_0} and
is a near-perfect liquid \cite{star_flow_2}.  PHENIX and STAR have
calculated the total charm cross-section $\sigma_{c\bar{c}}$ for
$p+p$, $d$ + Au, Cu + Cu, and Au + Au collisions at
$\sqrt{s_{NN}}=200$ GeV
\cite{phenix_nphe,star_charm_dAu,baumgart,phenix_nphe_AuAu,star_nphe}.
The cross-section scales with the number of binary nucleon-nucleon
collisions, an indication that charm quarks are indeed produced
through initial parton hard-scattering. PHENIX and STAR have
measured the nuclear modification factor $R_{AA}$ for light-flavor
hadrons \cite{phenix_eta} and for non-photonic $e^{\pm}$, which come
predominantly from heavy-flavor semileptonic decays
\cite{phenix_nphe,phenix_nphe_AuAu,star_nphe}. The suppression of
non-photonic $e^{\pm}$ in central Au + Au collisions is greater than
expected. It has been difficult for theoretical models
\cite{djordjevic_1,armesto,wicks,van_hees,adil} to describe the
suppression of light-flavor hadrons and non-photonic $e^{\pm}$
simultaneously.  Perturbative QCD calculations (FONLL)
\cite{cacciari} predict that the $b$-decay contribution to the
non-photonic $e^{\pm}$ becomes comparable to the $c$-decay
contribution at $p_{T}\sim5$ GeV/$c$. STAR measures the difference
in azimuth angle $\phi$ for pairs of non-photonic $e^{\pm}$ and
hadrons \cite{x_lin}.  This distribution is fit with PYTHIA
simulations of the expected $c$- and $b$-decay contributions to
determine their relative strengths; these data are consistent with
the FONLL predictions.

\begin{acknowledgments}
Thanks to J. W. Harris, J. Bielcik, H. Caines, and T. Ullrich for
all they have taught.  Thanks to the Heavy Flavor Working Group at
STAR for their advice in preparing the talk.  S. Salur and S.
Baumgart provided valuable help in the preparation of this paper.
\end{acknowledgments}

\bigskip 

\begin{thebibliography}{9}   
\bibitem{karsch} F. Karsch, \textit{Nucl. Phys. A} \textbf{698} (2002) 199c
\bibitem{kanaya} K. Kanaya, \textit{Nucl. Phys. A} \textbf{715} (2003) 233c
\bibitem{star_energy} J. Adams \textit{et al}. (STAR Collaboration), \textit{Phys. Rev. C} \textbf{70} (2004) 054907
\bibitem{blaziot} J.-P. Blaizot, \textit{Nucl. Phys. A} \textbf{661} (1999) 3c
\bibitem{glauber} R. J. Glauber, \textit{Lectures on Theoretical Physics, Vol. 1}, p. 315 (1959)
\bibitem{star_glauber} J. Adams \textit{et al}. (STAR Collaboration), \textit{Phys. Rev. C} \textbf{70} (2004) 054907
\bibitem{braun_0} P. Braun-Munzinger \textit{et al.}, \textit{Phys. Lett. B} \textbf{518} (2001) 41
\bibitem{star_flow_0} C. Adler \textit{et al}. (STAR Collaboration), \textit{Phys. Rev. Lett.} \textbf{90} (2003) 032301
\bibitem{star_flow_1} C. Adler \textit{et al}. (STAR Collaboration), \textit{Phys. Rev. Lett.} \textbf{87} (2001) 182301
\bibitem{star_flow_2} C. Adler \textit{et al}. (STAR Collaboration), \textit{Phys. Rev. Lett.} \textbf{89} (2002) 132301
\bibitem{star_jets_0} C. Adler \textit{et al}. (STAR Collaboration), \textit{Phys. Rev. Lett.}
\textbf{90} (2003) 032301
\bibitem{star_jets_1} C. Adler \textit{et al}. (STAR Collaboration), \textit{Phys. Rev. Lett.} \textbf{89} (2002) 202301
\bibitem{mustafa} M. G. Mustafa, \textit{Phys. Rev. C} \textbf{72} (2005) 014905
\bibitem{djordjevic_0} M. Djordjevic, \textit{Phys. Rev. C} \textbf{74} (2006) 064907
\bibitem{phenix_gamma} S. S. Adler \textit{et al}. (PHENIX Collaboration), \textit{Phys. Rev.
Lett.} \textbf{94} (2005) 232301
\bibitem{phenix_pi0_0} S. S. Adler \textit{et al}. (PHENIX Collaboration), \textit{Phys. Rev. Lett.} \textbf{91} (2003) 072301
\bibitem{phenix_pi0_1} S. S. Adler \textit{et al}. (PHENIX Collaboration), \textit{Phys. Rev. C} submitted, arXiv:nucl-ex/0611007v1 (2006)
\bibitem{phenix_eta} S. S. Adler \textit{et al}. (PHENIX Collaboration), \textit{Phys. Rev. C} \textbf{75} (2007) 024909
\bibitem{vitev_0} I. Vitev and M. Gyulassy, \textit{Phys. Rev. Lett.} \textbf{89} (2002) 252301
\bibitem{vitev_1} I. Vitev, \textit{J. Phys. G} 30 (2004) S791
\bibitem{gyulassy_0} M. Gyulassy and L. McLerran, \textit{Nucl. Phys. A} \textbf{750} (2005) 30-63
\bibitem{rhic} H. Hahn \textit{et al}., \textit{Nucl. Instr. and Meth. A} \textbf{499} (2003) 245
\bibitem{star_detector} K. H. Ackerman \textit{et al}. (STAR Collaboration), \textit{Nucl. Instr. and Meth. A} \textbf{499} (2003) 624-632
\bibitem{phenix_detector} K. Adcox \textit{et al}. (PHENIX Collaboration), \textit{Nucl. Instr. and Meth. A} \textbf{499} (2003) 469
\bibitem{z_lin} Z. Lin and M. Gyulassy, \textit{Phys. Rev. C} \textbf{51} (1995) 2177
\bibitem{matsui} T. Matsui and H. Satz, \textit{Phys. Lett. B} \textbf{178} (1986) 416
\bibitem{phenix_jpsi_0} A. Adare, \textit{et al}. (PHENIX Collaboration), arXiv:nucl-ex/0611020v1 (2006)
\bibitem{phenix_jpsi_1} S. S. Adler \textit{et al}. (PHENIX Collaboration), \textit{Phys. Rev. Lett.} \textbf{96} (2006) 012304
\bibitem{phenix_jpsi_2} R. Averbeck, \textit{J. Phys. G} \textbf{34} (2007) S567-S574
\bibitem{braun_1} P. Braun-Munzinger and J. Stachel, \textit{Phys. Lett. B} \textbf{490} (2000) 196
\bibitem{braun_2} P. Braun-Munzinger and J. Stachel, \textit{Nucl. Phys. A} \textbf{690} (2001) 619c
\bibitem{grandchamp} L. Grandchamp \textit{et al}., \textit{Phys. Rev. Lett.} \textbf{92} (2004) 212301
\bibitem{thews} R. L. Thews and M. L. Mangano, \textit{Phys. Rev. C} \textbf{73} (2006) 014904
\bibitem{andronic} A. Andronic \textit{et al}., arXiv:nucl-th/0701079v2 (2007)
\bibitem{kharzeev} Y. L. Dokshitzer and D. E. Kharzeev, \textit{Phys. Lett. B} \textbf{519} (2001) 199
\bibitem{star_charm_dAu} J. Adams \textit{et al}. (STAR Collaboration), \textit{Phys. Rev. Lett.} \textbf{94} (2005) 062301
\bibitem{pdg} S. Eidelman \textit{et al}. (Particle Data Group), \textit{Phys. Lett. B} \textbf{592} (2004) 1
\bibitem{zhang} H. Zhang for the STAR Collaboration, \textit{Nucl. Phys. A} \textbf{774} (2006) 701
\bibitem{zhong} C. Zhong for the STAR Collaboration, \textit{J. Phys. G} \textbf{34} (2007) S741
\bibitem{hijing} X.-N. Wang and M. Gyulassy, \textit{Phys. Rev. D} \textbf{44} (1991) 3501-3516
\bibitem{star_nphe} B. I. Abelev \textit{et al}. (STAR Collaboration), \textit{Phys. Rev. Lett.} \textbf{98} (2007) 192301
\bibitem{suaide_0} A. A. P. Suaide (for the STAR Collaboration), \textit{J. Phys. G} \textbf{30} (2004) S1179-S1182
\bibitem{phenix_nphe} A. Adare \textit{et al}. (PHENIX Collaboration), \textit{Phys. Rev. Lett.} \textbf{97} (2006) 252002
\bibitem{phenix_nphe_AuAu} A. Adare \textit{et al}. (PHENIX Collaboration), \textit{Phys. Rev. Lett.} \textbf{98} (2007) 172301
\bibitem{baumgart} S. Baumgart for the STAR Collaboration, arXiv:0709.4223v1 [nucl-ex] (2007)
\bibitem{frixone} S. Frixione \textit{et al}., arXiv:hep-ph/9702287v2 (1997)
\bibitem{cacciari} M. Cacciari \textit{et al}., \textit{Phys. Rev. Lett.} \textbf{95} (2005) 122001
\bibitem{vogt} R. Vogt, arXiv:0709.2531v1 [hep-ph] (2007)
\bibitem{suaide_1} A. A. P. Suaide, \textit{J. Phys. G} 34 (2007) S369
\bibitem{star_light} J. Adams \textit{et al}. (STAR Collaboration), \textit{Phys. Rev. Lett.} \textbf{91} (2003) 172302
\bibitem{djordjevic_1} M. Djordjevic \textit{et al}., \textit{Phys. Lett. B} \textbf{632} (2006) 81
\bibitem{armesto} N. Armesto \textit{et al}., \textit{Phys. Lett. B} \textbf{637} (2006) 362
\bibitem{wicks} S. Wicks \textit{et al}., \textit{Nucl. Phys. A} \textbf{784} (2007) 426-442
\bibitem{van_hees} H. van Hees \textit{et al}., \textit{Phys. Rev. C} \textbf{73} (2006) 034913
\bibitem{adil} A. Adil and I. Vitev, arXiv:hep-ph/0611109v2 (2006)
\bibitem{x_lin} X. Lin (for the STAR collaboration), \textit{J. Phys. G} \textbf{34} (2007) S821

\end{thebibliography}

\end{document}